\documentstyle[12pt,a4]{article}
\newcommand{\be}{\begin{equation}}
\newcommand{\ee}{\end{equation}}
\newcommand{\bea}{\begin{eqnarray}}
\newcommand{\ena}{\end{eqnarray}}
\newcommand{\vs}[1]{\rule[- #1 mm]{0mm}{#1 mm}}

\newcommand{\epem}{$\ e^+ \ e^-\ $}

\newcommand{\pT}{${p_{T} }\ $}

\newcommand{\alf}{\alpha}

\newcommand{\alfspi}{ {\alpha_s(\mu) \over 2 \pi} }
\newcommand{\alfspis}{ \left( {\alpha_s(\mu) \over 2 \pi} \right) ^2 }

\newcommand{\dsigd}{  {d \sigma^{D} \over {d {\vec p}_T d \eta}}}
\newcommand{\dsigsf}{{d \sigma^{SF} \over {d {\vec p}_T d \eta}}}
\newcommand{\dsigdf}{{d \sigma^{DF} \over {d {\vec p}_T d \eta}}}
\newcommand{\dsiggg}{ {d \sigma^{\gamma \gamma \rightarrow jet}
            \over {d {\vec p}_T d \eta}}}
\newcommand{\dsigig}{{d \sigma^{i \gamma \rightarrow jet}
            \over {d {\vec p}_T d \eta}}}

\newcommand{\dsiggj}{{d \sigma^{\gamma j \rightarrow jet}
            \over {d {\vec p}_T d \eta}}}
\newcommand{\dsigij}{{d \sigma^{i j \rightarrow jet}
            \over {d {\vec p}_T d \eta}}}

\newcommand{\NP}[1]{Nucl.\ Phys.\ {\bf #1}}
\newcommand{\PL}[1]{Phys.\ Lett.\ {\bf #1}}

\newcommand{\PR}[1]{Phys.\ Rev.\ {\bf #1}}

\newcommand{\ZPH}[1]{Z.\ Phys.\  {\bf #1}}

\begin{document}

\renewcommand{\thefootnote}{\fnsymbol{footnote}}
\newpage
\setcounter{page}{0}

%\vs{10}
\vs{9}

\begin{center}
{\Large {\bf{The production of jets from quasi real photons }}} \\
{\Large {\bf{ in \epem collisions}}} \\
%\vspace{1,5cm}
\vspace{0.7 cm}
{\large P. Aurenche, J.-Ph. Guillet} \\
{\em Laboratoire de Physique Th\'eorique ENS{\large{\em L}}APP
\footnote{URA 14-36 du
CNRS, associ\'ee \`a l'Ecole Normale Sup\'erieure de Lyon, et au
Laboratoire d'Annecy-le-Vieux de Physique des Particules.} $-$ Groupe
d'Annecy\\
LAPP, IN2P3-CNRS, B.P. 110, F-74941 Annecy-le-Vieux Cedex, France}
\\[0.7cm]
{\large M. Fontannaz} \\
{\em Laboratoire de Physique Th\'eorique et Hautes Energies
\footnote{ Laboratoire associ\'e au CNRS (URA 63). }
\\ Universit\'e de Paris XI, b\^atiment 211, F-91405 Orsay Cedex,
France}
\\[0.7cm]
{\large Y. Shimizu, J. Fujimoto} \\
{\em Physics Department, KEK \\
Tsukuba, Ibaraki 305, Japan} \\[0.8cm]
{\large K. Kato} \\
{\em Physics Department, Kogakuin University \\
Shinjuku, Tokyo 160, Japan} \\[0.8cm]
\end{center}
%\vs{7}
\vs{4}

\centerline{ \bf{Abstract}}
%\vs{5}
\vs{3}

We consider the production of jets in photon-photon collisions beyond
the leading logarithm approximation. Theoretical uncertainties as well
as uncertainties due to the virtuality of the initial photons are
discussed in detail. The comparison with TOPAZ data is performed and good
agreement is found between experiment and theory. It is expected that
future high precision TRISTAN data will constrain the non-perturbative
component of the photon structure function.

\vs{3}

\rightline{ENSLAPP-A-482/94}
\rightline{KEK preprint 94-66}
\rightline{KEK CP-019}
\rightline{July 1994}

%%%%%\vs{3}

%\renewcommand{\thefootnote}{\arabic{footnote}}
%\setcounter{footnote}{0}
\newpage

\section{Introduction}

\indent

The study of large transverse momentum processes in photon-photon
collisions already has a long history since in the late 70's and early
80's, several experimental collaborations at DESY and SLAC have
collected
data on the reaction $\gamma \gamma \rightarrow h \ X$ \cite{tas,mark}.
The most recent results concerning this process were published last
year \cite{mark} and are shown in Fig. 1
where the next-to-leading-logarithmic QCD predictions are
also displayed \cite{us}: the situation is
somewhat puzzling since "perfect" agreement between theory and data is
obtained at rather low transverse momentum where the theory is not very
reliable (because of the importance of the poorly known hadronic or
$VDM$ component of the photon) while the theory falls below the data, by a
factor 2 to 5, at large \pT where the physics is dominated by the "QED"
process $\gamma \gamma \rightarrow q \bar{q}$. We observe a disagreement
in the $p_T$ dependence which
may be attributed to the fact that the data follow the scaling behavior
$p^3_T \ d \sigma /dp_T = {\rm constant}$ \cite{mark},
whereas one would expect this quantity to
fall with \pT due to the $q \rightarrow h$ fragmentation process.

In the following we consider another version of large \pT processes,
namely $\gamma \gamma \rightarrow jet\  X$. We first present
the theoretical expressions and discuss some uncertainties inherent to
the perturbative approach. Since the data concern the reaction
$e^+ e^- \rightarrow e^+e^-\ jet \ X$ via two photon exchange we have
to study the validity of the Weizs\"acker-Williams \cite{wewi}
approximation and the effect of the photon virtuality.
Finally we compare the theory with the very recent TOPAZ anti-tag data
\cite{top,amy} and stress the importance of the gluon content of
the photon in the kinematical range covered by the data.

\section{Theoretical expressions for $\gamma + \gamma \rightarrow
jet +X$}

\indent

As extensively discussed at this  workshop \cite{sf}, the photon can
couple to the hard sub-process either directly or through its quark or
gluon content. The cross section for the production of a jet of a given
\pT and pseudorapidity $\eta$ can therefore be decomposed as \cite{gaga}
\be
\frac{d\sigma}{d\vec{p}_T d \eta} \ = \ \dsigd + \ \dsigsf + \ \dsigdf
\ee
where each term is now being specified. Beyond the leading logarithm
approximation the "direct" cross section takes the form
\be
\dsigd (R) =  \dsiggg \ + \ \alfspi K^{D} (R;M) .
\label{eq:dir}
\ee
with the corresponding diagrammatic decomposition shown in Fig.2.
The parameter $R$ specifies the jet cone size \cite{top}, while $\mu$
and $M$ are the renormalization and factorization scales respectively.
When one photon couples directly and the other one through its
"structure function", it leads to
\bea
\dsigsf (R) &=& \sum_{i=q,g} \int dx_{1} F_{i/\gamma}(x_{1},M)
\nonumber \\
&\ & \ \ \ \ \ \alfspi \left( \dsigig + \ \alfspi K^{SF}_{i \gamma}
(R;M, \mu) \right) \nonumber \\
 &+& \sum_{j=q,g} \int dx_{2} F_{j/\gamma}(x_{2},M)
 \nonumber \\
&\ & \ \ \ \ \ \alfspi \left( \dsiggj + \ \alfspi K^{SF}_{\gamma j}
(R;M, \mu) \right)
\label{eq:sf}
\ena
where some diagrams representative of the ${\cal{O}}(\alf_s)$ and
${\cal{O}} (\alf_s^2)$ terms on the right hand side are shown in
Fig. 3 a) and b) respectively. The underlined diagrams in Fig. 2 b)
and 3 a) are in fact the same but they contribute to different region
of phase space. When the final state quark is not collinear to
the initial photon (as in Fig. 2 b)) the exchanged propagator has
a large virtuality (shown by the fat line)  and the corresponding
contribution is associated to the \underline{hard} subprocess
$K^D$; when the final quark becomes almost collinear to the initial
photon (as in Fig. 3 a)) the virtuality of the exchanged propagator
is small: the interaction is \underline{soft} (long range) and
the corresponding contribution reflects the properties of the photon
fragmenting in a $q \bar{q}$ pair and is then naturally associated
to the photon structure function. Roughly speaking the factorization
scale $M$ separates the hard region from the soft region and
changing this arbitrary scale shifts contributions from $d\sigma^D$ to
$d\sigma^{SF}$ but clearly does not affect the sum $d\sigma^D + d
\sigma^{SF}$. More precisely, the photon structure function satisfies
the evolution equation of type \cite{wit}.
\be
{d F_{i/\gamma}(M) \over d \ln M^2} =   P_{i \gamma}
+ \sum_{j=q,g}P_{ij} \otimes F_{j/\gamma}(M)
\label{eq:evolsimp}
\ee
where the $P_{i\gamma},\ P_{ij}$ are the relevant Altarelli-Parisi
splitting functions. The scale variation associated to the
inhomogeneous term, $P_{i\gamma} $,
induces a change in $d \sigma^{SF}$ which is compensated by
a corresponding variation of $K^D(R,M)$ as described above: this effect
is unique to reactions involving photons as external legs. As for
the remaining variation associated to the homogeneous term in eq.(4)
it is compensated by a variation of $K^{SF} (R,M,\mu)$, as it occurs
in purely hadronic reactions.

We turn now to the last component in eq.(1) where both photons
interact via their structure functions
\bea
\dsigdf &=& \sum_{i,j=q,g} \int dx_{1} dx_{2}
 F_{i/\gamma}(x_{1},M) F_{j/\gamma}(x_{2},M) \alfspis \nonumber \\
&\ & \ \ \ \ \ \ \ \
\left( \dsigij \ + \ \alfspi K^{DF}_{ij} (R;M,\mu) \right)
\label{eq:df}
\ena
and the corresponding diagrams in Fig. 4 a) and b). Similarly to our
previous discussion, the higher order diagrams of $d\sigma^{SF}$
generate the "Born" contribution to $d\sigma^{DF}$ as well as
the higher order contributions $K^{SF}$ to $d\sigma^{SF}$ with
the consequence that scale dependent terms in $\sum_i
\left( \frac{\alf_s}{2\pi} \right)^2 F_{i\gamma} (M)
\otimes K^{SF}$ compensate the variation in $M$ of $ d \sigma^{DF}$.

In conclusion, only the sum eq.(1) has a physical meaning.
In particular, it is not legitimate to associate $d \sigma^{SF}$ and
$d \sigma^{DF}$ to experimentally measured "once resolved" and "twice
resolved" components. Let us finally remark that the renormalization
scale $\mu$ variation is compensated, as in purely hadronic cross
sections, within the "Born" and higher-order corrections in eqs. (3)
and (5) separately.

To illustrate quantitatively the variation of the theoretical
predictions under changes of $M$ and $\mu$ we consider jet production
at TRISTAN ($\sqrt{s}_{e^+e^-} = 58 \ GeV, \ p_T = 5.24 \ GeV/c$).
The photon structure functions of ref. \cite{font} have been used and
the proper convolutions have been made to construct, from the
$\gamma  \gamma \rightarrow jet \ X$ reaction, the $e^+ e^-$ cross
section with the relevant experimental cuts of TOPAZ \cite{top}
(see below). Fig. 5 a) is obtained when setting arbitrarily
$K^D = K^{SF} = K^{DF} =0$ (the so-called leading logarithmic
predictions) while Fig. 5 b) takes into account the full expressions
eqs. (2)-(5): the gain in stability is remarkable despite the fact
that no saddle-point or extremum is found \cite{gaga}.
The same patterns are also observed at $\sqrt{s}_{e^+e^-} = 1 \ TeV$.
In the following we always use for definiteness $M=\mu=p_T$.

\section{From $\gamma \gamma$ to \epem cross sections}

\indent

The usual procedure is to use the Weiz\"acker-Williams approach
\cite{top,kess} which approximates the \epem cross section by
the convolution
\bea
{d \sigma^{ee \rightarrow jet} \over {d {\vec p}_T d \eta}} =
\int dz_{1} dz_{2}  \ F_{\gamma/ e}(z_{1},E) \ F_{\gamma/ e}(z_{2},E)
{d \sigma^{\gamma \gamma \rightarrow jet } \over {d {\vec p}_T d \eta}}
\label{eq:equiv}
\ena
where $F_{\gamma/e}(z,E)$ is the spectrum of collinear photons emitted
by an electron or positron of energy $E$ (see Fig. 6). In the above
approximation one has neglected the dependence of the $\gamma \gamma$
cross section on the virtuality $q^2,\ {q'}^{2}$ of the photons.

Usually, experiments have an anti-tagging condition which restricts
the angle between the incoming and outgoing  electron
$\theta < \theta_{Max}$ such that  $q^2_{Max} = - E^2 (1-z)
\theta^2_{Max}$ for sufficiently small angles.
The quasi-real photon spectrum becomes
\bea
F_{\gamma/ e}(z,E,\theta_{Max})
         = {\alpha \over \pi} \left( {1+(1-z)^2 \over z} \right)
   \ln\ { E(1-z) \theta_{max} \over z\ m_e }
\label{eq:antitag}
\ena
with $m_e$, the electron mass. Neglecting the photon virtualities in
$d \sigma^{\gamma \gamma}$ typically introduces an error of
($\widehat{s}$ is the energy of the $\gamma \gamma$ system)
\[
\frac{|q^2|}{\hat{s}} \ \leq \ \frac{E^2 \theta^2_{Max}}{4p_T^2}
\]
which is less than $10\%$ in the case of TOPAZ
($\theta_{Max} = 3.2^\circ, \ p_T  \geq\ 2.5 \ GeV/c$). For large \pT
processes another effect may become relevant.  A photon of virtuality
$q^2$ has a transverse momentum given by $k^2_T \sim -(1-z)q^2$
which may be as high as $2.5 \ GeV^2$ in the case of TOPAZ.
This $k_T$ plays the role of an "intrinsic" transverse momentum
in hadronic collisions and it is well-known that, because of
"trigger bias" effects, this intrinsic momentum distorts the shape of
the jet \pT distribution at not too large \pT. Since the
Weisz\"acker-Williams spectrum (eq.(7)) assumes the photons
to be collinear to
the electron this effect is neglected in eq.(6). To quantitatively
test the reliability of our approximations we consider the process
$e^+e^- \rightarrow e^+ e^- \mu^+ \mu^-$
%%%%%%\cite{minam}
(same kinematics as $e^+e^- \rightarrow e^+ e^- q \bar{q})$
and study the distribution $\frac{d\sigma}{d\vec{p}_T d\eta}$
for a single $\mu$ using, on the one hand, the exact matrix element
and, on the other hand, eq.(6) with the photon spectrum eq.(7).
For the case of TOPAZ we find that the approximate result overestimates
the exact one by $8\%$, independent of \pT up to $p_T = 8.5 \ GeV/c$.
Such a correction will be included in our subsequent calculation.

This analysis covers the effects of the photon virtuality in the direct
process(eq.(2)) but, unfortunately, it is not complete in the case
of the $SF$ and $DF$ processes. Indeed, for these processes there
appear new terms, in the photon structure functions, of type
$\ln M^2/|q^2|$ or $q^2/m_\rho^2$  where $m_\rho$ is a typical vector
meson mass as we are now going to discuss.  Consider the shematic $SF
$ process as shown in Fig.7 where the parton $i$ hard
scatters with a characteristic mass scale $M$ to produce a jet.

In a schematic notation the cross section for the initial electron
to produce a jet can be written
\be
d\sigma^{e\rightarrow jet} = \frac{\alf}{2\pi} \int_0^1 \frac{dz}{z}
\ \frac{1+(1-z)^2}{z} \int^{Q^2_{Max}}_{Q^2_{Min}} \frac{dQ^2}{Q^2}
\int^z_0 dx
\ F_{i/\gamma} (\frac{x}{z}, M, Q) \ d \sigma^{i \rightarrow jet}
\ee
where we have introduced the structure function of a virtual $\gamma$
of mass $q^2 =-Q^2$, with the condition $Q^2<<M^2$. It satisfies eq.(4)
and similarly to the real photon case the solution to the evolution
equation is written as
\be
F_{i/\gamma} (z,M,Q) = F^{Pert}_{i/\gamma} (z,M,Q)
                                   + F_{i/\gamma}^{VDM} (z,M,Q).
\ee

We now make a model for each term on the right hand side.
Following the analysis of ref. \cite{bor} we assume
\be
\begin{array}{llll}
F^{Pert}_{i/\gamma} (z,M,Q) & \equiv & F^{Pert}_{i/\gamma} (z,M) \sim \ln
\frac{M^2}{Q_0^2} & \mbox{for} \ \ Q^2 < Q^2_0 \nonumber \\
F^{Pert}_{i/\gamma} (z,M,Q) &\sim & \ln \frac{M^2}{Q^2} &
                   \mbox{for} \ \ Q^2 \geq Q^2_0
\end{array}
\ee
where the function with two arguments on the right hand side of the equation
refers to the real photon. The value of $Q_0^2$ is chosen to be $.5 \ GeV^2$
as in ref. \cite{font}. Inserting this in eq.(8) we find
\bea
d\sigma^{e\rightarrow jet}|_{pert} \sim \left( \ln{ M^2 \over Q^2_0}
\ln{Q^2_{max} \over Q^2_{min}} - \frac{1}{2}
\ln^2{Q^2_{max} \over Q^2_0} \right) d\sigma^{i \rightarrow jet}
\label{eq:anotag}
\ena
The first term would be obtained, had we used $F_{i/\gamma} (z,M)$ over
the $Q^2$ integration range while the second term is the reduction
factor due to the $Q^2$ dependence of $F_{i/\gamma}$. For TOPAZ it
never exceeds a negligible $2\%$.
Following an observation of Borzumati and Schuler \cite{bor}, it is
argued in ref.\cite{dre} that the gluon structure function should be
more suppressed (as $\ln^2 \frac{M^2}{Q^2}$ when $Q^2 \geq Q^2_0$)
than the quark structure function and should lead to a
somewhat larger reduction factor in that case.

Turning to the $VDM$ component
in eq.(9) we make the usual $\rho$-pole dominance ansatz and write
\be
F^{VDM}_{i/\gamma} (z,M,Q) =
\left( \frac{m_\rho^2}{m_\rho^2 + Q^2} \right)^2
\ F^{VDM}_{i/\gamma} (z,M)
\ee
to obtain
\be
d\sigma^{e \rightarrow jet}|_{VDM} \sim \left( \ln \frac{Q^2_{Max}}{Q^2_{Min}}-
\left(\ln \frac{m^2_\rho+Q^2_{Max}}{m^2_\rho} +
\frac{Q^2_{Max}}{m^2_\rho+Q^2_{Max}}\right) \right) d\sigma^{i \rightarrow jet}
\ee
where the correction factor is now $10\%$ to $12\%$ in the case of
TOPAZ and is taken into account in our numerical estimates.

\section{Comparison to TOPAZ data and conclusions}

\indent

The TOPAZ collaboration has measured the single jet spectrum at
$\sqrt{s}_{e^+e^-} = 58 \ GeV$ under some specific anti-tagging
conditions (mainly $\theta <3.2^\circ$): the error bars include
the systematic errors added linearly to the statistical ones \cite{top}
(Fig. 8). The next-to-leading QCD predictions, based on the photon
structure function of ref.\cite{font}, are also shown in the
figure: the top curve is obtained using the standard set of structure
functions derived from a comparison to the $F_2^\gamma(x,M)$ deep
inelastic data \cite{dat} while the bottom curve is obtained
by arbitrarily setting the $VDM$ component equal to 0 in
eq.(9). Both curves are compatible with the data for $p_T > 4 \ GeV/c$.
The middle curve is the prediction when the $VDM$ component in
$F_{i/\gamma}(x,M,Q)$ is divided by 2 a choice still compatible with
the deep-inelastic photon data \cite{dat}. Little change is observed
in the predictions when instead of varying the $VDM$ normalization one
varies the shape of the quark and gluon distributions in
$F_{i/\gamma}^{VDM}$ \cite{gaga}.
Concerning the role of the higher order corrections, we find that
for the scales $M=\mu=p_T$ they increase the lower order result by
$25\%$ at $p_T = 3 \ GeV/c$ and leave it practically unchanged at
large \pT. The pattern of the higher order corrections is quite
different for the different components of eq.(1): while $d\sigma^D$ is
decreased by $15\%$, independently of \pT, and $d\sigma^{SF}$ is left
practically unchanged, $d\sigma^{DF}$ is increased by $70\%$.

In conclusion, we find it extremely encouraging that the theory is
able to account for both the data on the deep-inelastic photon
structure function and jet production in $\gamma \gamma$ collisions.
The somewhat too high theoretical predictions at low \pT using our
"standard" set of structure functions is attributed to the neglect,
in the calculation, of the  charm quark mass. An estimate of this
effect leads to a reduction of the cross section of about $15\%$
at $p_T =3\ GeV/c$ and only $2\%$ at large \pT. Taking this into account
the agreement of our standard set of predictions with the data is quite good.
It is obvious that the new TOPAZ (and AMY) data, with errors reduced by
a factor 2, will provide a very powerful tool to constrain
the non-perturbative input to the photon structure function.
Combining this with future results from LEP on photon deep-inelastic
scattering \cite{f2} as well as jet photoproduction
\cite{erd,exp} will lead \cite{jet} to a quantitative understanding
on the hadronic structure of the photon.

\medskip

\centerline{{\bf Acknowledgements}}

\indent

We would like to thank CNRS-IN2P3 (France) and Monbusho (Japan) for
the support to our collaboration. Three of us (P.A., J.-Ph.G., M.F)
are also indebted to the EEC programme "Human Capital and Mobility",
Network "Physics at High Energy Colliders", contract CHRX-CT93-0357 (DG 12
COMA) for financial support. They also thank Prof. G. Jarlskog for the
organization of a very interesting and enjoyable workshop.


\begin{thebibliography}{99}
%
\bibitem{tas}	TASSO collaboration: R.~Brandelik et al., \PL {B107} (1981)
		290; \PL {B138} (1984) 219.
\bibitem{mark}  MARK II collaboration: D.~Cords et al., \PL {B302} (1993) 341.
\bibitem{us}	P.~Aurenche, R.~Baier, A.~Douiri, M.~Fontannaz and D.~Schiff,
                \ZPH {C29} (1985) 423; CERN yellow report CERN 86-02 (1986)
		193, J. Ellis and R. Peccei eds;     \\
		D.J.~Miller, ECFA workshop on LEP 200, CERN 87-08 (1987) 202,
		A.~Bohm and W.~Hoogland eds.
\bibitem{wewi}  C.F.~Weizs\"acker, \ZPH {88} (1934) 612; \\
                E.J.~Williams, \PR {45} (1934) 729.
\bibitem{top}   TOPAZ collaboration: H.~Hayashii et al., \PL {B314} (1993) 149.
\bibitem{amy}   see also: AMY collaboration, R.~Tanaka et al.,
		\PL {B325} (1994) 248.
\bibitem{sf}	See in particular the talks by M.~Fontannaz, A.~Vogt and
  		P.~Zerwas at this workshop.
\bibitem{gaga}	P.~Aurenche, J.-Ph.~Guillet, M.~Fontannaz, Y.~Shimizu,
		J.~Fujimoto and K.~Kato, Prog. Theor. Phys. {\bf 92} (1994) 175.
\bibitem{wit}   E.~Witten, \NP {B120} (1977) 189; \\
		W.A.~Bardeen and A.J.~Buras, \PR {D20} (1979) 166; \PR {D21}
                (1980) 2041 E.
\bibitem{font}  M.~Fontannaz, Orsay preprint LPTHE-93-22, talk given at the
                21st International meeting on Fundamental Physics, Miraflores
                de la Sierra, Spain, May 1993; \\
                P.~Aurenche, M.~Fontannaz and J.Ph.~Guillet, LPTHE 93-37,
		October 1993.
\bibitem{kess}  see the talk by P.~Kessler at this workshop.
%%%%%\bibitem{minam} Minami-Tateya collaboration: could you please provide a
%%reference?
\bibitem{bor}	G.A.~Schuler, CERN-TH.6427/92,
                proceedings of the  DESY workshop on physics at HERA, Hamburg,
                Oct. 1991; \\
                F.M.~Borzumati and G.A.~Schuler, \ZPH {C58} (1993) 139.
\bibitem{dre}	M.~Drees and R.M.~Godbole, U. of Wisconsin preprint, MAD/PH/819,
		March 1994.
\bibitem{dat}   PLUTO collaboration: Ch.~Berger et al., \NP {B281} (1987) 365;
		\\
		AMY collaboration: T.~Sasaki et al., \PL {B252} (1990) 491; \\
		JADE collaboration: W.~Bartel et al., \ZPH {C24} (1984) 231.
\bibitem{f2}	F.~Kapusta, talk at this workshop; \\
		D.~Miller, $ibid$.
\bibitem{erd}	see the talk by M.~Erdmann at this workshop.
\bibitem{exp}	H1 collaboration: I.~Abt et al, \PL {B314} {1993} 436;  \\
		ZEUS collaboration: M.~Derrick, \PL {B322} (1994) 287.
\bibitem{jet}	D.~B\"odeker, G.~Kramer and S.G.~Salesh, DESY preprint DESY
		94-042; \\
		J.Ph.~Guillet, talk at this workshop; preprint in preparation.
%\bibitem{hera}  H1 collaboration: T.~Ahmed et al., \PL {B297} (1992) 205,
%                T.~Abt et al., \PL (B314) (1993) 436; \\
%		ZEUS collaboration: M.~Derrick et al., \PL {B316} (1993) 412.


\end{thebibliography}
\end{document}